%% file: arXiv_Draft.tex
\documentclass[letterpaper]{article} 
\usepackage[draft]{aaai2026}  
\usepackage{times}  
\usepackage{helvet}  
\usepackage{courier}  
\usepackage[hyphens]{url}  
\usepackage{graphicx} 
\usepackage{natbib}  
\usepackage{caption} 
\usepackage[utf8]{inputenc} 
\usepackage[T1]{fontenc}    
\usepackage{url}            
\usepackage{booktabs}       
\usepackage{amsfonts}       
\usepackage{nicefrac}       
\usepackage{microtype}      
\usepackage[most]{tcolorbox}  %
\usepackage{listings}         %
\usepackage{xcolor,colortbl}
\usepackage{subcaption}

\usepackage{graphicx}
\usepackage{mathtools}
\usepackage{booktabs}
\usepackage{makecell}

\newcommand{\eg}{\textit{e.g.}}

\definecolor{mypink}{RGB}{255, 143, 171}
\definecolor{mygreen}{RGB}{86, 171, 145}
\usepackage{multirow}
\usepackage{enumitem}

\newcommand{\datasetName}{\texttt{P$^2$V}}
\newcommand{\datasetNameClean}{\texttt{P$^2$V[o]}}
\newcommand{\datasetNameNoised}{\texttt{P$^2$V[p]}}

\newcommand{\nPublicFigures}{206~}

\newcommand{\nContext}{10~}

\definecolor{DPLightGreen}{rgb}{0.307, 0.610, 0.446}
\definecolor{DPLightBlue}{rgb}{0.541, 0.585, 0.951}
\definecolor{DPLightRed}{rgb}{0.941, 0.585, 0.551}
\definecolor{DPLightOrange}{RGB}{147, 140, 122}

\definecolor{RBRed}{rgb}{0.98,0.88,0.85}
\definecolor{RBRedLight}{rgb}{1,0.93,0.90}

\definecolor{RBBlue}{rgb}{0.81,0.80,0.94}
\definecolor{RBBlueLight}{rgb}{0.91,0.90,0.98}

\definecolor{RBBlueL}{rgb}{0.84,0.83,0.97}
\definecolor{RBBlueLL}{rgb}{0.86,0.85,0.99}

\definecolor{RBGreen}{rgb}{0.85,0.91,0.84}
\definecolor{RBGreenLight}{rgb}{0.93,0.98,0.92}

\definecolor{RBOrange}{RGB}{250,240,210}

\definecolor{DPBlueD}{rgb}{0.24,0.43,0.77}
\definecolor{DPPurple}{rgb}{0.46,0.12,0.45}

\definecolor{DPGreen}{rgb}{0,0.45,0.24}
\definecolor{DPLightGreen}{rgb}{0,0.65,0.44}

\tcbset{
  promptbox/.style={
    colback=blue!10,         
    colframe=blue!60,         
    fonttitle=\bfseries,    %
    title=Prompt,       %
    boxrule=0.8pt,          %
    arc=2mm,                %
    left=4pt, right=4pt, top=4pt, bottom=4pt
  }
}
\tcbset{
  promptboxICwC/.style={
    colback=blue!10,         
    colframe=blue!60,         
    fonttitle=\bfseries,    %
    title=ICwC Prompt,       %
    boxrule=0.8pt,          %
    arc=2mm,                %
    left=4pt, right=4pt, top=4pt, bottom=4pt
  }
}
\tcbset{
  promptboxICwoC/.style={
    colback=blue!10,         
    colframe=blue!60,         
    fonttitle=\bfseries,    %
    title=ICwoC Prompt,       %
    boxrule=0.8pt,          %
    arc=2mm,                %
    left=4pt, right=4pt, top=4pt, bottom=4pt
  }
}
\tcbset{
  promptboxOCwC/.style={
    colback=blue!10,         
    colframe=blue!60,         
    fonttitle=\bfseries,    %
    title=OCwC Prompt,       %
    boxrule=0.8pt,          %
    arc=2mm,                %
    left=4pt, right=4pt, top=4pt, bottom=4pt
  }
}
\tcbset{
  promptboxOCwoC/.style={
    colback=blue!10,         
    colframe=blue!60,         
    fonttitle=\bfseries,    %
    title=OCwoC Prompt,       %
    boxrule=0.8pt,          %
    arc=2mm,                %
    left=4pt, right=4pt, top=4pt, bottom=4pt
  }
}

\usepackage{algorithm}
\usepackage{algorithmic}

\usepackage{newfloat}
\usepackage{listings}
\DeclareCaptionStyle{ruled}{labelfont=normalfont,labelsep=colon,strut=off} 
\floatstyle{ruled}
\newfloat{listing}{tb}{lst}{}
\floatname{listing}{Listing}
\title{Perturbed Public Voices (\texttt{P$^{2}$V}): A Dataset for Robust Audio Deepfake Detection}
\author{
    Chongyang Gao\textsuperscript{\rm 1}\equalcontrib, Marco Postiglione\textsuperscript{\rm 1}\equalcontrib,
    Isabel Gortner\textsuperscript{\rm 1}, Sarit Kraus\textsuperscript{\rm 2}, V.S. Subrahmanian\textsuperscript{\rm 1}
}
\affiliations{
    \textsuperscript{\rm 1}Northwestern University, Evanston, Illinois, USA\\
    \textsuperscript{\rm 2}Bar-Ilan University, Ramat Gan, Israel\\
    vss@northwestern.edu 
}

\begin{document}

\maketitle

\begin{abstract}
Current audio deepfake detectors cannot be trusted. While they excel on controlled benchmarks, they fail when tested in the real world. We introduce Perturbed Public Voices (\datasetName), an IRB-approved dataset capturing three critical aspects of malicious deepfakes: (1) identity-consistent transcripts via LLMs, (2) environmental and adversarial noise, and (3) state-of-the-art voice cloning (2020–2025). Experiments reveal alarming vulnerabilities of 22 recent audio deepfake detectors: models trained on current datasets lose 43\% performance when tested on \datasetName, with performance measured as the mean of F1 score on deepfake audio, AUC, and 1-EER. Simple adversarial perturbations induce up to 16\% performance degradation, while advanced cloning techniques reduce detectability by 20-30\%. In contrast, \datasetName-trained models maintain robustness against these attacks while generalizing to existing datasets, establishing a new benchmark for robust audio deepfake detection. \datasetName~will be publicly released upon acceptance by a conference/journal. 
\end{abstract}


\input{sections/introduction}

\input{sections/related_work}

\input{sections/dataset}

\input{sections/experiments}

\section{Limitations}

\datasetName\ currently focuses on English-speaking deceased public figures and does not include multilingual or live speaker data. Similar to most audio deepfake detection benchmarks (e.g. ASVspoof \cite{todisco2019asvspoof}, WaveFake \cite{frank2wavefake}, Codecfake \cite{xie2024codecfake}), \datasetName\ contains a disproportionately high share of deepfake audio compared to real. This skew diverges from current real‑world settings where real speech outnumbers deepfake content. We recommend researchers and practitioners to take this into account to ensure not overestimating deepfake detection capabilities in realistic scenarios.

\section{Conclusion}

Audio deepfake detection faces a critical trust gap: current systems excel on artificial benchmarks but fail against real-world challenges. In light of this, we have proposed a novel large-scale and IRB-approved dataset, namely Perturbed Public Voices (\datasetName), consisting of 257,440 samples from 206 deceased public figures and incorporating news-guided transcripts as well as stateof-the-art audio generation techniques and perturbations that can be encountered in the real world. We have demonstrated that adversarial perturbations and evolving synthesis methods systematically degrade state-of-the-art detectors performance, exposing vulnerabilities that existing datasets overlook. Our results establish that robustness requires training on data that mirrors the complexity of malicious deepfakes, including noise, plausible content, and state-of-the-art voice cloning. \datasetName~provides this foundation, enabling models that maintain reliability against attacks while generalizing to in-the-wild scenarios.

\bibliography{aaai2026}

\appendix

\input{sections/appendix}

\end{document}

%% file: sections/introduction.tex
\section{Introduction}


Rapid advances in generative AI have ushered in an era of hyper-realistic synthetic media, with audio deepfakes emerging as a powerful tool for misinformation \cite{DBLP:conf/aaai/ZhangYWZZ024,DBLP:conf/aaai/ChenYF0RZZYGXWL25}. From impersonating public figures in fraudulent speeches\footnote{\url{https://www.npr.org/2024/05/30/nx-s1-4986088/deepfake-audio-elections-politics-ai}} to manipulating financial markets with fabricated CEO statements\footnote{\url{https://blackbird.ai/blog/celebrity-deepfake-narrative-attacks}}, malicious applications of this technology threaten democratic discourse, security, and trust in digital content \cite{DBLP:conf/aaai/WalkerSS24,DBLP:conf/icwsm/RuffinSX024}. While researchers have developed benchmark datasets and detectors~\cite{yamagishi2021asvspoof,yi2023audio,DBLP:conf/aaai/ZhangHL0G25}, existing resources fail to capture the complexity of real-world deepfakes \cite{muller2022does,wu2024clad}, where synthetic audio is often embedded in noisy environments, semantically aligned with a speaker’s identity, and generated using SOTA voice cloning methods.

Current datasets, such as those derived from the ASVspoof Challenge series~\citep{kinnunen2018automatic,wu2017asvspoof,kinnunen2017asvspoof,todisco2019asvspoof,yamagishi2021asvspoof,wang2024asvspoof}, prioritize controlled laboratory conditions, omitting critical variables like environmental noise and contextual plausibility. Recent efforts like In-The-Wild~\citep{muller2022does} and DEEP-VOICE~\citep{bird2023real} compile online deepfakes of celebrities but have limited speaker diversity and generalization ability as they rely on easily detected deepfakes that lack the sophistication of real-world malicious deployments. This mismatch between training data and real-world conditions creates a false sense of security: \textit{detection models achieve near-perfect performance on benchmark tasks but fail when confronted with the nuanced, adversarially perturbed deepfakes in the real world.}

We present Perturbed Public Voices (\datasetName), a large-scale, IRB-approved\footnote{IRB approval number: STU00223255} dataset that bridges this gap. It consists of 257,440 samples (247,200 fake, 10,240 real) from 206 deceased public figures. By systematically integrating transcripts tailored to a speaker's public persona using 3 different Large Language Models (LLMs), acoustic diversity with 10 perturbations simulating natural and adversarial distortions, and 10 state-of-the-art text-to-speech/voice cloning and voice conversion methods (spanning 2020-2025), we enable the development of detection systems robust to emerging threats. 

Our comprehensive experiments, involving 22 state-of-the-art audio deepfake detection methods, reveal stark limitations in current approaches: models trained on existing benchmarks show dramatic performance drops when tested on \datasetName---e.g., deepfake detection scores, measured as the mean of F1 scores on the fake class, AUC and 1-EER, decaying by 43\%---, while \datasetName-trained models generalize effectively to in-the-wild data. Additionally, simple adversarial perturbations (e.g., background noise and Gaussian noise injection) can cause up to 16\% performance degradation of deepfake detectors. Beyond benchmarking, \datasetName~provides a framework for evaluating the detectability of novel synthesis techniques. We demonstrate that newer voice cloning methods (e.g., XTTS-v2~\citep{casanova2024xtts} and IndexTTS~\citep{deng2025indextts}) are 20–30\% harder to detect than older approaches. By open-sourcing \datasetName, we aim to catalyze progress toward trustworthy audio authentication.

%% file: sections/related_work.tex
\section{Related Work}

Early efforts to establish publicly available benchmarks for audio deepfake detection, such as the ASVspoof Challenge series~\citep{kinnunen2018automatic,todisco2019asvspoof,yamagishi2021asvspoof,wang2024asvspoof}, established standard protocols for evaluating deepfake detectors. These datasets, grounded in laboratory conditions, pair clean speech recordings with synthetic audio produced via classical text-to-speech (TTS) and voice conversion (VC) models. While they support algorithmic benchmarking, they omit key factors present in real-world settings, including environmental noise, adversarial perturbations, and identity-context alignment.

Subsequent initiatives sought to increase ecological validity. The Audio Deep Synthesis Detection (ADD) Challenge~\citep{yi2022add,yi2023add} introduced background noise and music, while other datasets embedded short fake segments within real utterances~\citep{zhang2021initial,zhang2022partialspoof}. Recent datasets have extended scope along multiple axes: multilinguality (e.g., WaveFake~\citep{frank2wavefake}, CFAD~\citep{ma2024cfad}, MLAAD~\citep{muller2024mlaad}), emotional manipulation (e.g., EmoFake~\citep{zhao2024emofake}), domain-specific scenarios (e.g., Fake Song Detection~\citep{xie2024fsd}), or detailed method annotations (e.g., SFR~\citep{yan2022system}, Codecfake~\citep{xie2024codecfake}). TIMIT-TTS~\citep{salvi2023timit} incorporates visual cues for multimodal detection, while CLAD~\citep{wu2024clad} explores semantic editing.

The In-The-Wild (ITW) dataset~\citep{muller2022does} compiles deepfakes of 58 public figures sourced from online platforms. However, many samples are easily identified as fake, as they feature well-known personalities speaking absurd or fictional content (e.g., Donald Trump reads Star Wars). As the authors note, these clips are suitable for evaluation, but inadequate for training, due to their semantic implausibility and limited diversity. Similarly, the DEEP-VOICE dataset~\citep{bird2023real} focuses on celebrities but lacks consistent annotation and perturbation modeling.

\datasetName\ addresses key limitations in existing corpora by systematically varying both synthesis and acoustic conditions. First, it incorporates deepfakes generated using ten state-of-the-art TTS and VC models spanning diverse architectures and training paradigms, enabling a controlled evaluation of how different generative techniques impact detection difficulty. Second, each deepfake is based on semantically aligned transcripts derived from real public discourse by well-known figures, ensuring contextual plausibility and speaker consistency. Third, to emulate real-world deepfakes, we introduce ten perturbations (e.g., Gaussian noise, background noise, air absorption) that reflect both environmental variability and potential adversarial interference. Audio samples are annotated with the corresponding generation method and perturbation, allowing for fine-grained attribution of detection performance across synthesis pipelines and noise conditions.

%% file: sections/dataset.tex
\section{Perturbed Public Voices (\datasetName)}

\begin{figure*}[t]
\begin{center}
\centerline{\includegraphics[width=\textwidth]{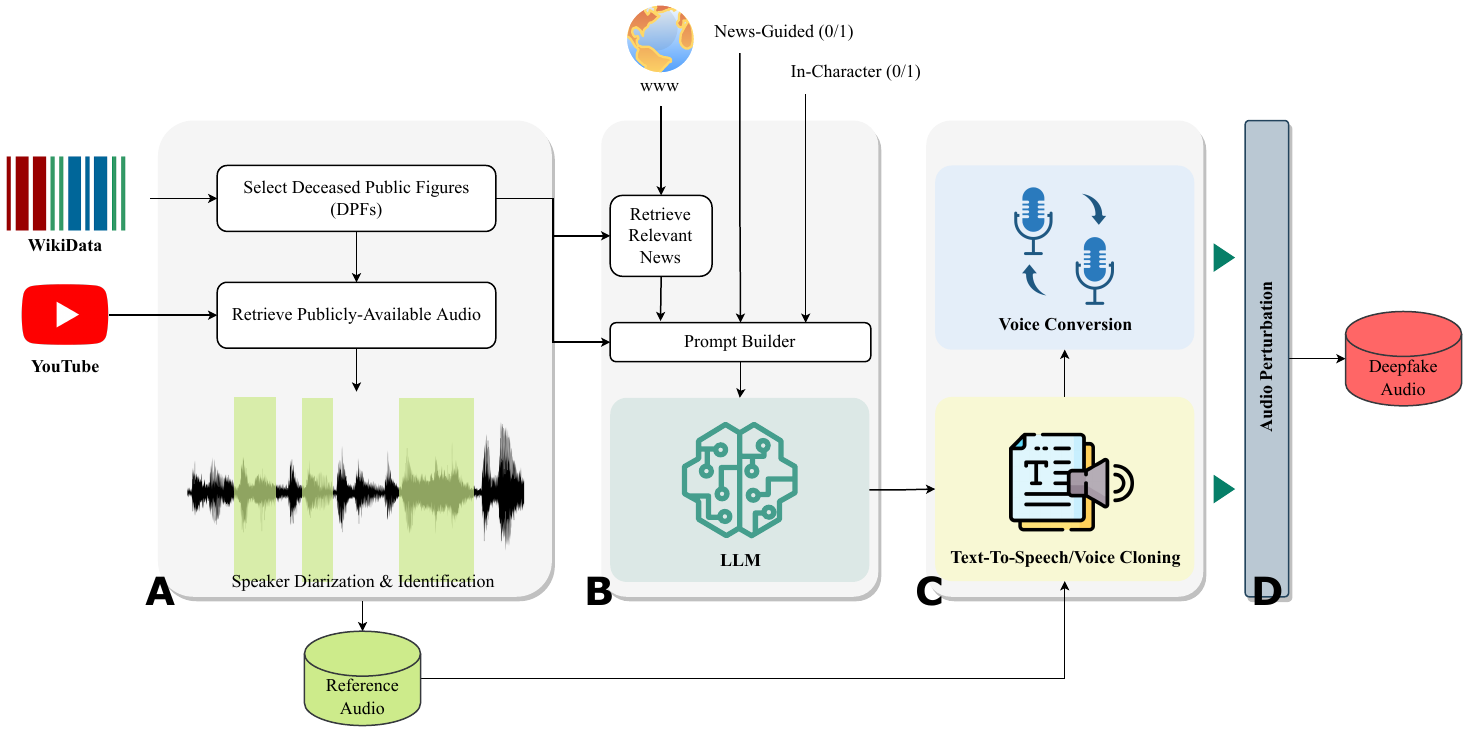}}
\caption{\textbf{Overview of the audio deepfake generation pipeline for the \datasetName~dataset.} The workflow consists of four main stages: (A) \textit{Reference audio collection} from WikiData and YouTube, followed by speaker diarization and identification to isolate target speaker segments; (B) \textit{Fake transcript generation} using Large Language Models (LLMs) with news-guided and in-character/out-of-character conditioning; (C) \textit{Audio deepfake synthesis} through Text-to-Speech (TTS) models and Voice Conversion (VC) methods; and (D) \textit{Audio perturbation application} to simulate real-world acoustic conditions and adversarial distortions, ultimately producing the final deepfake audio dataset.}
\label{fig:overview}
\end{center}
\end{figure*}

This section provides a detailed description of how we build our dataset. The overview of our dataset construction workflow is illustrated in Figure~\ref{fig:overview}.

\subsection{Reference Audio Collection}

\paragraph{Selection of Deceased Public Figures (DPFs)}
We identified deceased public figures (DPFs) for the construction of \datasetName~using WikiData. Our data collection spanned from January 1, 2012, to May 1, 2025, focusing on individuals from professions susceptible to deepfake manipulation: politicians, journalists, actors, musicians, writers, television presenters, film directors, philosophers, and activists. We conducted manual curation to verify the availability of high-quality audio samples for \nPublicFigures public figures.

\paragraph{Publicly-Available Audio Retrieval} 
To simulate realistic conditions under which malicious actors might operate (i.e., accessing public sources to obtain reference speech for voice cloning of public figures), we collected authentic audio samples from YouTube. Our search methodology used keywords specifically targeting interviews, speeches, panel discussions, podcasts, and other public speaking engagements featuring the identified public figures. We prioritized high-quality recordings with minimal background noise, clear articulation, and extended duration to ensure sufficient speech material for robust voice analysis. This approach mirrors the practical constraints and opportunities that would confront those attempting unauthorized voice replication, thereby enhancing the ecological validity of our dataset.

\paragraph{Speech Diarization and Speaker Identification}
Since the collected audio samples for the \nPublicFigures DPFs contain multiple speakers, we isolate the target DPFs' speech content. We used Pyannote \cite{Plaquet23,Bredin23}, a state-of-the-art speech diarization framework that segments audio into discrete units (corresponding to individual utterances) and assigns unique speaker IDs to each segment. Following diarization, we implemented a post-processing step where consecutive segments with identical speaker identifiers were merged into single contiguous segments, with a maximum duration of 40 seconds. This automated process was also manually verified to accurately associate speaker IDs with either the target DPF or an \textit{unknown} classification. To minimize privacy concerns, segments associated with \textit{unknown} speakers were excluded from the dataset. For quality control and consistency, we also excluded segments shorter than 5 seconds (which typically contain insufficient reference content for voice cloning methods) and longer than 60 seconds (which may include unwanted artifacts or speaker transitions due to diarization errors). 

\subsection{Fake Speech Transcript Generation}

To simulate a broad spectrum of real-world scenarios in which synthetic audio might be deployed, we enriched our dataset with temporally grounded transcripts associated with each DPF. Specifically, we leveraged the date of death for each individual to construct a timeline of potentially relevant public discourse. For each DPF, we selected \nContext random distinct articles published within the five years preceding their death. These articles were identified through targeted search queries designed to retrieve publicly available content about the DPF at various points in time.

LLMs are increasingly demonstrating impressive capabilities in role-playing tasks~\citep{shanahan2023role,lu2024large}, including a nuanced understanding of public figures and celebrities~\citep{yokoyama2024aggregating,chen2024using,chen2024oscars}. Leveraging this strength, we generate four distinct categories of transcripts by combining: in-character versus out-of-character speech patterns, and news-guided versus non-news-guided content. 

We prioritized models that balance performance and computational efficiency, specifically those with 10 to 30 billion parameters. To ensure transparency and reproducibility, we use open-source models rather than proprietary ones such as ChatGPT or Gemini. We selected three instruction-tuned models from established open-source LLM families that consistently demonstrated strong instruction adherence and formatting compliance~\citep{ouyang2022training,wei2021finetuned}: \emph{Gemma-3}~\citep{team2025gemma} with 27B parameters, \emph{Qwen2.5}~\citep{yang2025qwen2} with 14B parameters, and \emph{Mistral}~\footnote{\url{https://huggingface.co/mistralai/Mistral-Small-24B-Instruct-2501}} with 24B parameters. To ensure consistency in audio duration (approximately 30 seconds), each LLM was prompted to generate transcripts containing around 100 words. For each model, we use the same prompts (see Appendix). 



\subsection{Audio Deepfake Generation}

To generate high-fidelity audio deepfakes, we combined our curated transcripts with public reference audio using ten state-of-the-art generative pipelines, comprising seven text-to-speech (TTS) models and three voice conversion (VC) methods. Each TTS model synthesizes speech conditioned on a semantically aligned transcript and a randomly selected target-speaker reference. We employ diverse architectures, including Tacotron DCA~\citep{battenberg2020location}, YourTTS~\citep{casanova2022yourtts}, XTTS~\citep{casanova2024xtts}, VEVO~\citep{zhang2025vevo}, Zonos~\footnote{\url{https://github.com/Zyphra/Zonos}}, CosyVoice 2~\citep{du2024cosyvoice}, and IndexTTS~\citep{deng2025indextts}, covering a range of capabilities from multilingual and zero-shot synthesis to fine-grained control over prosody and emotional expression. To increase speaker diversity and evaluate layered synthesis effects, we additionally adopt a two-stage pipeline in which speech generated by IndexTTS is transformed via voice conversion using Diff-HierVC~\citep{choi2023diff}, FreeVC~\citep{li2023freevc}, or Seed-VC~\citep{liu2024zero}. These models enable text-free, style-controllable, and speaker-adaptive conversion in zero-shot settings. All methods use officially released pretrained weights to ensure reproducibility. Each audio sample is labeled with its generative method, facilitating attribution in downstream experiments.

\subsection{Audio Perturbations}
To simulate real-world acoustic variability and assess model robustness, we apply one of ten perturbations to each generated audio deepfake, selected uniformly at random, with associated hyperparameters sampled per transformation. Each sample is labeled with its perturbation type and parameters to ensure full reproducibility. 

The ten perturbations are: (1) \textit{Background noise addition}: one audio file is sampled from the ESC-50 dataset~\citep{piczak2015esc}, which includes 2,000 labeled environmental sounds. (2) \textit{Gaussian noise injection}: noise amplitude is fixed at a minimum of 0.001 and sampled from {0.005, 0.01, 0.015, 0.02} for the maximum. (3) \textit{Air absorption simulation}: source–microphone distance is selected from {10, 20, 50, 100} meters to replicate frequency-dependent attenuation over distance. (4) \textit{MP3 compression}: lossy encoding is applied with bitrates drawn from {8, 16, 32, 64} kbps to reflect online transmission degradation. (5) \textit{Pitch shifting}: semitone offsets are sampled from ±2, ±4, ±6, and ±12 to test resilience to pitch-based manipulation. (6) \textit{Band-pass filtering}: frequency ranges are chosen from (200, 4000), (150, 5000), or (50, 8000) Hz, simulating telephone-quality or bandwidth-limited audio. (7) \textit{Time stretching}: playback speed is modified by a factor selected from {0.6, 0.8, 1.2, 1.4}, preserving pitch but altering rhythm and cadence. (8) \textit{Audio reversal}: the waveform is reversed in time, eliminating temporal structure without requiring parameter tuning. (9) \textit{Impulse response convolution}: reverberation is added using one audio sample randomly selected from the EchoThief Impulse Response Library\footnote{\url{https://www.echothief.com/}}. (10) \textit{Clipping distortion}: amplitude saturation is introduced using percentile thresholds drawn from {(0, 20), (10, 40), (20, 60)}, controlling distortion severity. These perturbations form a controlled yet varied augmentation suite designed to reflect environmental noise, adversarial conditions, and common signal degradations, thereby supporting rigorous and ecologically valid evaluation of audio deepfake detection systems.

\subsection{Dataset Statistics and Organization} 
\label{sec:organization}

Table~\ref{tab:dataset} provides a breakdown of the dataset, illustrating the systematic organization of samples across different generation methods and data splits. The dataset contains 257,040 audio samples derived from 206 unique subjects, partitioned into training (131 subjects), validation (37 subjects), and test (38 subjects) sets to ensure robust model evaluation and prevent data leakage across speaker identities. The dataset organization follows standard practices ~\citep{sun2023ai}, with approximately 60\% of samples allocated to training, 20\% to validation, and 20\% to testing. The subject-level partitioning guarantees that no speaker appears across multiple splits, thereby preventing overfitting to specific vocal characteristics and promoting generalization to unseen speakers. Like most existing benchmarks (e.g. ASVspoof \cite{todisco2019asvspoof}, WaveFake \cite{frank2wavefake}, CodecFake \cite{xie2024codecfake}), \datasetName\ contains a disproportionately high number of fake audio, in contrast to current real-world scenarios where real speech prevails. To address potential biases, we report class-specific metrics in our experiments and encourage future research to carefully take this imbalance into account.

To facilitate comprehensive evaluation under varying acoustic conditions, we provide both a clean, original version of the dataset \datasetName\texttt{[o]}, and a perturbed version \datasetName\texttt{[p]} incorporating noise perturbations.

\begin{table}[t]
\centering
\caption{\datasetName\ distribution of subjects and audio samples across training, validation, and test splits. The dataset contains both synthetic (fake) audio generated through various methods and authentic (real) audio samples.}
\label{tab:dataset}
\small
\scalebox{0.74}{
\begin{tabular}{@{}l|ccc|c@{}}
\toprule
\textbf{Category} & \textbf{Training} & \textbf{Validation} & \textbf{Test} & \textbf{Total} \\ 
\midrule 
Subjects & 131 & 37 & 38 & 206 \\ 
\midrule
\multicolumn{5}{c}{\textit{Synthetic Audio Samples}} \\
\midrule
Samples per LLM (3) & 52,400 & 14,800 & 15,200 & 82,400\\
Samples per Voice Cloning Method (10) & 15,720 & 4,440 & 4,560 & 24,720 \\ 
Samples per Noise Perturbation (10) & 15,720 & 4,440 & 4,560 & 24,720 \\
\textbf{Fake Audios (Total)} & \textbf{157,200} & \textbf{44,400} & \textbf{45,600} & \textbf{247,200} \\
\midrule
\textbf{Real Audios} & \textbf{6,237} & \textbf{1,963} & \textbf{2,040} & \textbf{10,240} \\
\midrule
\textbf{Grand Total} & \textbf{163,437} & \textbf{46,363} & \textbf{47,640} & \textbf{257,440} \\
\bottomrule
\end{tabular}}
\end{table}

%% file: sections/experiments.tex
\section{Experiments}

Our experiments examine four key dimensions: (1) cross-dataset generalization performance comparing our dataset against existing benchmarks (InTheWild \citep{muller2022does}), (2) the impact of state-of-the-art voice cloning technologies incorporated in our dataset on detection accuracy, (3) the influence of diverse transcript generation methods on detection systems, and (4) vulnerability to common audio perturbations that adversaries might employ to evade detection. These analyses reveal significant limitations in current detectors and validate our dataset's utility in highlighting the evolving challenges posed by advancing synthesis technologies.

\subsection{Experimental Setup}
\subsubsection{Dataset}

We demonstrate the robustness and generalization ability of \datasetName\ against a related real-world dataset, In-The-Wild (ITW)~\cite{muller2022does}. The ITW dataset contains deepfakes of 58 public figures extracted from publicly available video and audio content, containing 19,963 authentic and 11,816 deepfake samples. Similarly to \datasetName, ITW was partitioned into training, validation and test sets using stratified sampling to preserve the original class distribution.

The ITW dataset consists of speakers engaging in absurd and out-of-character dialogue, which may introduce biases that could limit model generalization when trained exclusively on such data. To investigate this aspect, we conduct extensive cross-dataset evaluations between \datasetName\ and ITW. Given the class imbalance differences between \datasetName\ and ITW (with ITW having a higher proportion of authentic samples), we report comprehensive class-specific performance metrics including precision, recall, and F1-scores for both authentic and deepfake classes.

\subsubsection{Detection Models}
Following~\citet{kawa2023improved}, we adopt two types of audio deepfake detection pipelines. The first is an end-to-end architecture, where the model directly processes raw audio input and is trained in a fully end-to-end manner. This category includes RawNet3~\citep{jung2022pushing}. The second type is a feature-based pipeline, where audio features are first extracted (Linear Frequency Cepstral Coefficients (LFCC)~\citep{davis1980comparison}, Mel-Frequency Cepstral Coefficients (MFCC)~\citep{sahidullah2012design}, and the Whisper
features~\cite{radford2023robust}) and then passed to a classifier for analysis. Representative models in this category include LCNN \citep{wu2018light}, MesoNet \citep{afchar2018mesonet}, and SpecRNet \citep{kawa2022specrnet}. 

These 4 model architectures and 3 feature types (with their 7 combinations) are used to train a set of 22 baseline audio deepfake detectors. These models output a binary prediction indicating the likelihood of the audio being a deepfake. Models are trained using the binary cross-entropy loss, defined as:
\begin{equation}
\label{equ:le}
\mathcal{L} = -{(y \cdot log(p) + (1 - y)log(1 - p))}
\end{equation}
where $y$ in Equation \ref{equ:le} denotes the ground-truth label of the input audio and $p$ is the model's predicted probability.

Following the training settings in~\cite{kawa2023improved}, we use the Adam optimizer across all models. For RawNet3, the learning rate is set to $1e-3$ with a weight decay of $5e-5$. LCNN, MesoNet, and SpecRNet are trained using a learning rate of $1e-4$ and a weight decay of $1e-4$. We use a batch size of 16 and evaluate model performance in 5 epochs, as the model converges. Each experiment is repeated using random seeds 0, 1, and 2. All experiments are conducted using three NVIDIA RTX A6000 GPUs.

For brevity and clarity, results presented in this section represent ensemble predictions averaged across the 22 state-of-the-art baseline models. Detailed performance metrics for individual baselines on \datasetNameClean, \datasetNameNoised\ and ITW are reported in the Appendix (Tables~\ref{tab:cfa}, ~\ref{tab:cfa_noised} and ~\ref{tab:itw}).

\subsubsection{Evaluation Metrics}
The evaluation metrics include precision, recall, and F1 scores for each class individually. We also compute overall precision and recall using weighted averages, where the number of instances per class determines the weights. In addition, we report AUC and Equal Error Rate (EER), a widely used metric in deepfake detection that captures the balance point between false acceptance and false rejection rates. We compute an aggregated Deepfake Detection Score (DDS) to provide a concise summary of the model’s overall performance:
\begin{equation}
    DDS = (F1_{Fake} + AUC + (1-EER))/3
\end{equation}
For ease of interpretation, all reported metrics (except for EER) are scaled by a factor of 100. 

\subsection{Results}

\subsubsection{Cross-Dataset Evaluations}
\label{sec:cross-dataset}

\begin{figure}
    \centering
    \includegraphics[width=\linewidth]{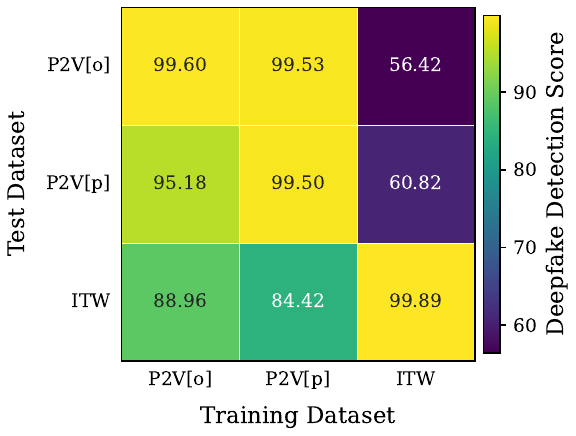}
    \caption{\textbf{Cross-dataset generalization performance across audio deepfake detection datasets.} Each cell reports the DDS performance (mean of AUC, F1 score on the fake class, and 1–EER) obtained by averaging predictions across 22 baseline models trained on the dataset indicated on the x-axis and evaluated on the dataset indicated on the y-axis.}
    \label{fig:cross-dataset-averaged}
\end{figure}

We assessed the generalization ability of models trained on \datasetName---both in its original \datasetNameClean\ and perturbed \datasetNameNoised\ form---as well as on the ITW dataset. Figure~\ref{fig:cross-dataset-averaged} reports DDS scores averaging predictions across 22 baseline models, with rows indicating the test dataset and columns the training dataset. Results for all the metrics are reported in the Appendix. 

Across all configurations, models performed best when evaluated on data from the same distribution as their training set. The ITW-trained models achieved near-perfect performance when tested on ITW (99.89), but their performance dropped markedly when applied to either \datasetNameClean\ (56.42) or \datasetNameNoised\ (60.82)---demonstrating limited out-of-domain transfer and a high degree of dataset specificity. Interestingly, ITW-trained models performed better on \datasetNameNoised\ than on \datasetNameClean. This counterintuitive result likely occurs because ITW's real-world audio naturally contains noise patterns similar to those we artificially introduced in our perturbed dataset.

In contrast, models trained on \datasetNameNoised\ showed markedly improved cross-domain robustness. These models generalized well to \datasetNameClean\ (99.6 vs. 99.53 for models trained directly on the original) and maintained solid performance when evaluated on ITW (84.42). While models trained on \datasetNameClean\ achieved slightly higher performance on ITW (88.96), the difference was not statistically significant (paired t-test, $p = 0.39$).

The ITW performance degradation highlights the tendency of current detection systems to overfit to dataset-specific characteristics rather than learning generalizable synthetic speech patterns. The superior cross-domain performance of \datasetNameNoised-trained models suggests that exposure to diverse acoustic conditions during training enhances robustness to real-world deployment scenarios.

\subsubsection{Impact of TTS and VC Methods}
\label{sec:impact-voice-cloning}

\begin{table*}[t]
\centering

\caption{\textbf{Impact of TTS and VC Methods.} Performance comparison of predictions averaged across 22 deepfake detection models trained on ITW and evaluated on audio generation methods contained in \datasetNameNoised. Each row presents results obtained by filtering the test data to include only samples from the corresponding audio generation method. Metrics include precision (P), recall (R), and F1-score for both real and fake audio detection, along with overall AUC, Equal Error Rate (EER), and Deepfake Detection Score (DDS). Methods are ordered by publication year. Green shading in the DDS column indicates performance above baseline (60.82\%), while red shading indicates performance below baseline.}
\label{tab:impact-voice-cloning}
\scalebox{0.85}{
\small
\renewcommand{\arraystretch}{1} 
\begin{tabular}{@{}r|ccc|ccc|ccccc|c@{}}
\toprule
 &
  \multicolumn{3}{c}{\cellcolor{RBGreen}Real} &
  \multicolumn{3}{c}{\cellcolor{RBRed}Fake} &
  \multicolumn{6}{c}{\cellcolor{RBBlue}Overall} \\ \cmidrule(l){2-13} 
\textbf{Audio Generation Model} & 
  P $(\uparrow)$ &
  R $(\uparrow)$ &
  F1 $(\uparrow)$ &
  P $(\uparrow)$ &
  R $(\uparrow)$ &
  F1 $(\uparrow)$ &
  P $(\uparrow)$ &
  R $(\uparrow)$ &
  F1 $(\uparrow)$ &
  AUC $(\uparrow)$ &
  EER  $(\downarrow)$ & 
  DDS $(\uparrow)$ \\ \midrule
\rowcolor{gray!20} 
\textit{Baseline} & 4.85 & 99.90 & 9.25 & 99.96 & 12.31 & 21.92 & 52.41 & 56.11 & 15.58 & 84.54 & 0.240 & 60.82  \\ \midrule 
Tacotron2 DCA~\citep{battenberg2020location} & 33.59 & 99.90 & 50.28 & 99.62 & 11.64 & 20.85 & 66.61 & 55.77 & 35.56 & 94.49 & 0.131 & \cellcolor{RBGreen}67.40 \\
YourTTS~\cite{casanova2022yourtts} & 36.12 & 99.90 & 53.06 & 99.79 & 20.96 & 34.65 & 67.96 & 60.43 & 43.85 & 92.24 & 0.158 & \cellcolor{RBGreen} 70.37 \\
Diff-HierVC~\citep{choi2023diff} & 34.42 & 99.90 & 51.20 & 99.71 & 14.85 & 25.84 & 67.06 & 57.37 & 38.52 & 85.14 & 0.234 & \cellcolor{RBGreen} 62.54 \\
FreeVC~\cite{li2023freevc} & 32.95 & 99.90 & 49.55 & 99.52 & 9.04 & 16.57 & 66.23 & 54.47 & 33.06 & 91.58 & 0.175 & \cellcolor{RBGreen} 63.55 \\
CosyVoice2~\citep{du2024cosyvoice} & 34.13 & 99.90 & 50.88 & 99.68 & 13.75 & 24.17 & 66.91 & 56.83 & 37.52 & 81.53 & 0.272 & \cellcolor{RBRed} 59.51 \\
XTTS-v2~\citep{casanova2024xtts} & 32.01 & 99.90 & 48.49 & 99.15 & 5.09 & 9.68 & 65.58 & 52.49 & 29.08 & 74.11 & 0.322 & \cellcolor{RBRed} 50.52 \\
Seed-VC~\cite{liu2024zero} & 32.90 & 99.90 & 49.50 & 99.51 & 8.84 & 16.23 & 66.20 & 54.37 & 32.86 & 76.38 & 0.305 & \cellcolor{RBRed} 54.05 \\
Zonos (2025) & 35.84 & 99.90 & 52.76 & 99.78 & 20.00 & 33.32 & 67.81 & 59.95 & 43.04 & 93.11 & 0.144 & \cellcolor{RBGreen} 70.67 \\
IndexTTS~\citep{deng2025indextts} & 32.70 & 99.90 & 49.27 & 99.46 & 8.00 & 14.82 & 66.08 & 53.95 & 32.04 & 74.79 & 0.320 & \cellcolor{RBRed} 52.52 \\
VEVO~\citep{zhang2025vevo} & 33.41 & 99.90 & 50.07 & 99.60 & 10.92 & 19.68 & 66.50 & 55.41 & 34.88 & 82.06 & 0.262 & \cellcolor{RBRed} 58.51 \\

\bottomrule
  
\end{tabular}%
}
\renewcommand{\arraystretch}{1} 
\end{table*}

We evaluate the performance of predictions averaged across 22 deepfake detection models trained on the ITW dataset across voice samples generated by TTS and VC methods. Table~\ref{tab:impact-voice-cloning} summarizes the results, where models are tested on \datasetNameNoised.

The baseline performance (computed on the full \datasetName\texttt{[p]} test set) achieves AUC of 84.54\% and DDS of 60.82\%. However, performance varies significantly across different voice cloning methods. We observe DDS scores ranging from 50.52\% (XTTS-v2) to 70.67\% (Zenos), indicating substantial differences in detection difficulty.

We identify a clear temporal pattern: older voice cloning methods exhibit higher detectability compared to recent approaches. Methods published in 2020-2022 achieve high DDS scores (Tacotron2 DCA: 67.40\%, YourTTS: 70.37\%), while most methods from 2024-2025 show reduced detectability (XTTS-v2: 50.52\%, Seed-VC: 54.05\%, VEVO: 52.52\%, IndexTTS: 52.52\%). Zenos (70.67\% DDS score) is the most detectable method, potentially due to artifacts in their generation processes that current detectors can exploit. Performance degradation on newer methods indicates limited generalization of detection models trained on older synthesis techniques, highlighting the need for adaptive detection strategies that can handle new generation capabilities.

To statistically validate performance differences, we conduct Mann-Whitney U tests with False Discovery Rate (FDR) correction for multiple comparisons. For each sample, we calculate the detection error as the absolute difference between the averaged detector prediction ($y_p$) and the ground truth label ($y$). Each voice cloning method's distribution of these detection errors is compared against all others. All pairwise comparisons yield statistically significant differences ($p < 0.001$ after FDR correction).

\subsubsection{Impact of Transcript Generation}

\datasetName\ incorporates transcripts generated by three different Large Language Models: Gemma-3 (27B parameters) \citep{team2025gemma}, Qwen2.5 (14B parameters) \citep{yang2025qwen2}, and Mistral (24B parameters)\footnote{https://huggingface.co/mistralai/Mistral-Small-24B-Instruct-2501}. We investigate whether these LLMs introduce artifacts that help deepfake detection systems.

Table~\ref{tab:impact-llm} presents the performance comparison of 22 deepfake detection models across samples containing transcripts generated by each LLM. The results show minimal variation in detection performance: Gemma-3 achieves an Avg score of 60.60\%, Qwen2.5 reaches 61.40\%, and Mistral attains 60.49\%. All three methods perform within a narrow range around the baseline (60.82\%).

To validate the statistical significance of observed differences, we conduct Mann-Whitney U tests with FDR correction for multiple comparisons. Each LLM's detection error distribution is compared against all others. The analysis reveals no statistically significant differences between transcript generation methods (Gemma-3: $p=0.48$, Mistral: $p=0.48$, Qwen2.5: $p=0.87$ after FDR correction). These findings indicate that current detection systems are not exploiting LLM-specific artifacts in transcript generation, and that transcript quality variations between these models do not create detectable patterns that influence voice cloning detection accuracy.

\begin{table*}[t]
\centering
\caption{\textbf{Impact of Transcript Generation Methods.} Performance comparison of predictions averaged across 22 deepfake detection models trained on ITW and evaluated on transcript generation methods (LLMs) contained in \datasetNameNoised. Each row presents results obtained by filtering the test data to include only samples generated with the corresponding LLM. Metrics include precision (P), recall (R), and F1-score for both real and fake audio detection, along with overall AUC, Equal Error Rate (EER), and Deepfake Detection Score (DDS). Green shading in the DDS column indicates performance above baseline (60.82\%), while red shading indicates performance below baseline.}
\label{tab:impact-llm}
\scalebox{0.97}{
\small
\renewcommand{\arraystretch}{1} 
\begin{tabular}{@{}r|ccc|ccc|ccccc|c@{}}
\toprule
 &
  \multicolumn{3}{c}{\cellcolor{RBGreen}Real} &
  \multicolumn{3}{c}{\cellcolor{RBRed}Fake} &
  \multicolumn{6}{c}{\cellcolor{RBBlue}Overall} \\ \cmidrule(l){2-13} 
\textbf{LLM} & 
  P $(\uparrow)$ &
  R $(\uparrow)$ &
  F1 $(\uparrow)$ &
  P $(\uparrow)$ &
  R $(\uparrow)$ &
  F1 $(\uparrow)$ &
  P $(\uparrow)$ &
  R $(\uparrow)$ &
  F1 $(\uparrow)$ &
  AUC $(\uparrow)$ &
  EER  $(\downarrow)$ & 
  DDS $(\uparrow)$ \\ \midrule
\rowcolor{gray!20} 
\textit{Baseline} & 4.85 & 99.90 & 9.25 & 99.96 & 12.31 & 21.92 & 52.41 & 56.11 & 15.58 & 84.54 & 0.240 & 60.82 \\ \midrule
Gemma-3 (27B) & 13.28 & 99.90 & 23.44 & 99.89 & 12.45 & 22.14 & 56.59 & 56.17 & 22.79 & 84.03 & 0.244 & \cellcolor{RBRed} 60.60 \\
Qwen2.5 (14B) & 13.27 & 99.90 & 23.43 & 99.89 & 12.39 & 22.05 & 56.58 & 56.15 & 22.74 & 85.35 & 0.232 & \cellcolor{RBGreen}61.40 \\
Mistral (24B) & 13.23 & 99.90 & 23.37 & 99.89 & 12.09 & 21.56 & 56.56 & 55.99 & 22.47 & 84.25 & 0.243 & \cellcolor{RBRed} 60.49 \\

\bottomrule
  
\end{tabular}%
}
\renewcommand{\arraystretch}{1} 
\end{table*}

\subsubsection{Impact of Noise Perturbations}

Not only can malicious actors employ novel audio generation methods, but they can also use post-processing techniques to evade detection systems. We analyze the vulnerability of deepfake detectors to ten common audio perturbations that adversaries can easily apply to synthetic speech. Table~\ref{tab:impact-noised} presents the detection performance when samples are filtered by perturbation type, revealing significant variations in evasion effectiveness.

\begin{table*}[t]
\centering
\caption{\textbf{Impact of Noise Perturbations.} Performance comparison of predictions averaged across 22 deepfake detection models trained on ITW and evaluated on noise perturbations contained in \datasetNameNoised. Each row presents results obtained by filtering the test data to include only samples generated with the corresponding perturbation. Metrics include precision (P), recall (R), and F1-score for both real and fake audio detection, along with overall AUC, Equal Error Rate (EER), and Deepfake Detection Score (DDS). Green shading in the DDS column indicates performance above baseline (60.82\%), while red shading indicates performance below baseline.}
\label{tab:impact-noised}
\scalebox{0.87}{
\small
\renewcommand{\arraystretch}{1} 
\begin{tabular}{@{}r|ccc|ccc|ccccc|c@{}}
\toprule
 &
  \multicolumn{3}{c}{\cellcolor{RBGreen}Real} &
  \multicolumn{3}{c}{\cellcolor{RBRed}Fake} &
  \multicolumn{6}{c}{\cellcolor{RBBlue}Overall} \\ \cmidrule(l){2-13} 
\textbf{Noise Perturbation} & 
  P $(\uparrow)$ &
  R $(\uparrow)$ &
  F1 $(\uparrow)$ &
  P $(\uparrow)$ &
  R $(\uparrow)$ &
  F1 $(\uparrow)$ &
  P $(\uparrow)$ &
  R $(\uparrow)$ &
  F1 $(\uparrow)$ &
  AUC $(\uparrow)$ &
  EER  $(\downarrow)$ & 
  DDS $(\uparrow)$ \\ \midrule
\rowcolor{gray!20} 
\textit{Baseline} & 4.85 & 99.90 & 9.25 & 99.96 & 12.31 & 21.92 & 52.41 & 56.11 & 15.58 & 84.54 & 0.240 & 60.82 \\ \midrule
Audio Reversal & 32.33 & 99.90 & 48.86 & 99.35 & 6.69 & 12.54 & 65.84 & 53.30 & 30.70 & 92.09 & 0.159 & \cellcolor{RBGreen} 62.90 \\
Background Noise Addition & 30.68 & 99.90 & 46.94 & 96.88 & 1.33 & 2.62 & 63.78 & 50.62 & 24.78 & 68.45 & 0.363 & \cellcolor{RBRed} 44.92 \\
Impulse Response Application & 32.23 & 99.90 & 48.73 & 99.24 & 5.74 & 10.85 & 65.73 & 52.82 & 29.79 & 83.45 & 0.247 & \cellcolor{RBRed} 56.55 \\
Time Stretching & 32.97 & 99.90 & 49.58 & 99.46 & 8.18 & 15.11 & 66.22 & 54.04 & 32.35 & 84.01 & 0.248 & \cellcolor{RBRed} 58.11 \\
ClippingDistortion & 31.54 & 99.90 & 47.95 & 97.92 & 2.08 & 4.08 & 64.73 & 50.99 & 26.01 & 81.98 & 0.256 & \cellcolor{RBRed} 53.48 \\
Pitch Shifting & 33.26 & 99.90 & 49.91 & 99.56 & 9.91 & 18.03 & 66.41 & 54.91 & 33.97 & 86.02 & 0.227 & \cellcolor{RBRed} 60.45 \\
Band-Pass Filtering & 48.35 & 99.90 & 65.16 & 99.92 & 52.43 & 68.77 & 74.13 & 76.16 & 66.97 & 96.59 & 0.101 & \cellcolor{RBGreen} 85.08 \\
Air Absorption Simulation & 33.03 & 99.90 & 49.64 & 99.52 & 9.16 & 16.78 & 66.27 & 54.53 & 33.21 & 85.81 & 0.227 & \cellcolor{RBRed} 59.95 \\
MP3 Compression & 37.80 & 99.90 & 54.85 & 99.84 & 26.97 & 42.46 & 68.82 & 63.43 & 48.66 & 94.36 & 0.132 & \cellcolor{RBGreen} 74.53 \\
Gaussian Noise Injection & 31.11 & 99.90 & 47.45 & 89.47 & 0.38 & 0.75 & 60.29 & 50.14 & 24.10 & 72.82 & 0.331 & \cellcolor{RBRed} 46.82 \\
\bottomrule
  
\end{tabular}%
}
\renewcommand{\arraystretch}{1} 
\end{table*}

The results demonstrate a clear dichotomy in perturbation impact. Signal processing methods that preserve fundamental audio characteristics show minimal performance differences: Audio Reversal (62.90\%) and Impulse Response Application (56.53\%) remain close to the baseline (60.82\%). Similarly, traditional audio effects like Pitch Shifting (60.45\%) and Clipping/Distortion (53.48\%) show moderate impact on detection performance.

Conversely, frequency-domain and compression-based perturbations exhibit substantial effects on detection performance. Band-Pass Filtering achieves unexpectedly high detection accuracy (85.08\%), suggesting that frequency filtering may actually enhance detection by removing noise while preserving discriminative synthetic artifacts. MP3 Compression (74.53\%) similarly improves performance, likely due to the web-sourced nature of the InTheWild training dataset which contains compressed audio samples.

The most effective adversarial manipulations are Background Noise Addition (44.92\%) and Gaussian Noise Injection (46.82\%), providing attackers with approximately 15-point F1-score reductions from baseline. These techniques achieve extremely low fake detection F1-scores (2.75\% and 0.75\%, respectively) and high EERs (0.363 and 0.331, respectively), demonstrating that readily accessible and human-imperceptible noise injection can severely compromise detection systems.

To validate observed performance differences, we conduct Mann-Whitney U tests with False Discovery Rate (FDR) correction for multiple comparisons. Each noise perturbation method's detection error distribution is compared against all others. All pairwise comparisons yield statistically significant differences ($p < 0.001$ after FDR correction), confirming that performance variations are not due to random variation.



%% file: sections/appendix.tex
\section{Appendix}
\subsection{Transcript Generation Prompt}
\label{app:prompt}

This section details the prompt templates used to generate the four types of speech transcripts for our audio deepfake dataset. The prompts are designed to systematically vary two key dimensions: contextual grounding (news-guided vs. non-news-guided) and persona authenticity (in-character vs. out-of-character). The template is provided as follows:

\begin{tcolorbox}[promptbox]
\footnotesize 
if \textbf{News-Guided} is \textbf{True}:

\begin{verbatim}
  Consider this news piece along  with its 
  summary:\n"""\nTitle: {title_tokens} \n
  Summary: {summary_tokens}\n\"\"\"
\end{verbatim}
\begin{verbatim} 
Generate a realistic quote that would {not}
likely be said by {subject\_tokens},
maintaining their authentic speech 
patterns and vocabulary. 
The quote should be 100 words in 
length. Only output quote 
without other words.
\end{verbatim}
\end{tcolorbox}

Here, we are including the first section only when producing news-guided transcripts. In such cases, we provide the headline of the retrieved news article (\texttt{{title\_tokens}}) and a concise summary of the news article content (\texttt{summary\_tokens}). \texttt{subject\_tokens} represents the name of the deceased public figure (DPF) being impersonated, and \texttt{{not}} is a conditional modifier that is included for out-of-character conditions and omitted for in-character conditions.

\subsection{Detailed Experiments Results}
We presented the detailed results in this section across the 22 state-of-the-art baseline models. We train and test the baseline models on the same dataset. The detailed performance metrics for individual baselines trained and tested on \datasetNameClean, \datasetNameNoised\ and ITW are provided in Tables ~\ref{tab:cfa}, ~\ref{tab:cfa_noised} and ~\ref{tab:itw}.

\begin{table*}
\centering

\caption{Performance comparison of baselines on \datasetNameClean
with different feature subsets shown in parentheses. The DDS column is the average of F1-score on the fake class, AUC, and (1-EER). }
\label{tab:cfa}
\scalebox{0.91}{
\small
\renewcommand{\arraystretch}{1} 
\begin{tabular}{@{}l|ccccccccccc|cc@{}}
\toprule
 &
  \multicolumn{3}{c}{\cellcolor{RBGreen}Real} &
  \multicolumn{3}{c}{\cellcolor{RBRed}Fake} &
  \multicolumn{6}{c}{\cellcolor{RBBlue}Overall} & \\ \cmidrule(l){3-14} 
\textbf{Model} & 
  P $(\uparrow)$ &
  R $(\uparrow)$ &
  F1 $(\uparrow)$ &
  P $(\uparrow)$ &
  R $(\uparrow)$ &
  F1 $(\uparrow)$ &
  P $(\uparrow)$ &
  R $(\uparrow)$ &
  F1 $(\uparrow)$ &
  AUC $(\uparrow)$ &
  EER  $(\downarrow)$ & 
  DDS $(\uparrow)$ \\ \midrule

rawnet3 & 84.70 & 89.36 & 86.94 & 99.52 & 99.28 & 99.40 & 98.89 & 98.85 & 98.87 & 99.22 & 0.03 & 98.60 \\ 

 \midrule

(lfcc)-lcnn & 79.23 & 71.58 & 74.85 & 98.74 & 99.13 & 98.93 & 97.90 & 97.95 & 97.90 & 98.84 & 0.05 & 97.70 \\ 

(mfcc)-lcnn & 80.80 & 76.44 & 78.54 & 98.95 & 99.18 & 99.07 & 98.17 & 98.21 & 98.19 & 98.04 & 0.08 & 96.49 \\ 

(lfcc-mfcc)-lcnn & 75.81 & 73.50 & 74.49 & 98.82 & 98.93 & 98.87 & 97.83 & 97.84 & 97.83 & 96.73 & 0.10 & 95.28 \\ 

(whisper)-lcnn & 92.68 & 83.81 & 87.93 & 99.28 & 99.69 & 99.48 & 99.00 & 99.01 & 98.99 & 97.58 & 0.05 & 97.24 \\ 

(whisper-lfcc)-lcnn & 91.75 & 87.76 & 89.63 & 99.45 & 99.64 & 99.55 & 99.12 & 99.13 & 99.12 & 99.19 & 0.03 & 98.67 \\ 

(whisper-mfcc)-lcnn & 91.66 & 87.16 & 89.19 & 99.43 & 99.63 & 99.53 & 99.09 & 99.10 & 99.09 & 98.85 & 0.04 & 98.27 \\ 

(whisper-lfcc-mfcc)-lcnn & 90.89 & 85.85 & 88.20 & 99.37 & 99.61 & 99.49 & 99.01 & 99.02 & 99.01 & 99.01 & 0.04 & 98.24 \\ 
 \midrule

(lfcc)-mesonet & 5.57 & 99.62 & 10.54 & 99.95 & 22.46 & 34.55 & 95.91 & 25.77 & 33.52 & 93.03 & 0.12 & 71.79 \\ 

(mfcc)-mesonet & 54.77 & 18.01 & 11.08 & 96.35 & 95.35 & 95.71 & 94.57 & 92.04 & 92.09 & 86.88 & 0.22 & 86.83 \\ 

(lfcc-mfcc)-mesonet & 7.32 & 88.56 & 13.39 & 99.21 & 41.91 & 52.04 & 95.27 & 43.90 & 50.38 & 87.77 & 0.21 & 72.90 \\ 

(whisper)-mesonet & 11.72 & 64.41 & 18.26 & 98.46 & 65.47 & 67.93 & 94.75 & 65.43 & 65.80 & 95.08 & 0.11 & 84.15 \\ 

(whisper-lfcc)-mesonet & 38.88 & 61.39 & 41.59 & 98.26 & 92.48 & 95.06 & 95.71 & 91.15 & 92.77 & 93.05 & 0.14 & 91.40 \\ 

(whisper-mfcc)-mesonet & 64.91 & 48.73 & 31.78 & 97.66 & 88.30 & 91.80 & 96.25 & 86.61 & 89.23 & 90.60 & 0.17 & 88.57 \\ 

(whisper-lfcc-mfcc)-mesonet & 17.36 & 57.17 & 26.61 & 98.09 & 92.58 & 95.12 & 94.64 & 91.06 & 92.19 & 89.25 & 0.17 & 89.27 \\ 
 \midrule

(lfcc)-specrnet & 87.94 & 81.50 & 84.39 & 99.18 & 99.49 & 99.33 & 98.69 & 98.72 & 98.69 & 99.15 & 0.04 & 98.05 \\ 

(mfcc)-specrnet & 82.71 & 75.03 & 78.34 & 98.89 & 99.27 & 99.08 & 98.20 & 98.24 & 98.19 & 98.15 & 0.07 & 96.78 \\ 

(lfcc-mfcc)-specrnet & 81.68 & 79.59 & 79.86 & 99.09 & 99.16 & 99.13 & 98.35 & 98.33 & 98.30 & 97.86 & 0.06 & 97.00 \\ 

(whisper)-specrnet & 79.38 & 89.41 & 83.94 & 99.52 & 98.93 & 99.22 & 98.66 & 98.52 & 98.57 & 98.80 & 0.05 & 97.72 \\ 

(whisper-lfcc)-specrnet & 94.57 & 89.05 & 91.56 & 99.51 & 99.75 & 99.63 & 99.30 & 99.30 & 99.29 & 99.58 & 0.03 & 98.83 \\ 

(whisper-mfcc)-specrnet & 91.63 & 89.89 & 90.75 & 99.55 & 99.63 & 99.59 & 99.21 & 99.21 & 99.21 & 99.39 & 0.03 & 98.54 \\ 

(whisper-lfcc-mfcc)-specrnet & 97.06 & 87.06 & 91.78 & 99.42 & 99.88 & 99.65 & 99.32 & 99.33 & 99.31 & 99.43 & 0.03 & 98.68 \\ 

\bottomrule

\end{tabular}%
}
\renewcommand{\arraystretch}{1} 
\end{table*}

\begin{table*}
\centering

\caption{Performance comparison of baselines on \datasetNameNoised
with different feature subsets shown in parentheses. The DDS column is the average of F1-score on the fake class, AUC, and (1-EER). }
\label{tab:cfa_noised}
\scalebox{0.91}{
\small
\renewcommand{\arraystretch}{1} 
\begin{tabular}{@{}l|ccccccccccc|cc@{}}
\toprule
 &
  \multicolumn{3}{c}{\cellcolor{RBGreen}Real} &
  \multicolumn{3}{c}{\cellcolor{RBRed}Fake} &
  \multicolumn{6}{c}{\cellcolor{RBBlue}Overall} & \\ \cmidrule(l){3-14} 
\textbf{Model} & 
  P $(\uparrow)$ &
  R $(\uparrow)$ &
  F1 $(\uparrow)$ &
  P $(\uparrow)$ &
  R $(\uparrow)$ &
  F1 $(\uparrow)$ &
  P $(\uparrow)$ &
  R $(\uparrow)$ &
  F1 $(\uparrow)$ &
  AUC $(\uparrow)$ &
  EER  $(\downarrow)$ & 
  DDS $(\uparrow)$ \\ \midrule

rawnet3 & 90.28 & 66.37 & 76.50 & 98.51 & 99.68 & 99.09 & 98.16 & 98.25 & 98.13 & 96.56 & 0.09 & 95.65 \\
\midrule

(lfcc)-lcnn & 89.03 & 58.64 & 70.55 & 98.18 & 99.68 & 98.92 & 97.79 & 97.92 & 97.71 & 97.31 & 0.08 & 96.00 \\

(mfcc)-lcnn & 88.62 & 59.51 & 70.97 & 98.22 & 99.66 & 98.93 & 97.81 & 97.94 & 97.73 & 96.81 & 0.09 & 95.59 \\

(lfcc-mfcc)-lcnn & 79.33 & 68.87 & 73.70 & 98.62 & 99.19 & 98.90 & 97.79 & 97.89 & 97.82 & 97.32 & 0.09 & 95.80 \\

(whisper)-lcnn & 93.95 & 87.78 & 90.75 & 99.45 & 99.75 & 99.60 & 99.22 & 99.23 & 99.22 & 97.47 & 0.05 & 97.42 \\

(whisper-lfcc)-lcnn & 96.69 & 85.87 & 90.93 & 99.37 & 99.87 & 99.62 & 99.26 & 99.27 & 99.25 & 99.37 & 0.02 & 98.84 \\

(whisper-mfcc)-lcnn & 98.18 & 82.65 & 89.71 & 99.23 & 99.93 & 99.58 & 99.18 & 99.19 & 99.16 & 98.93 & 0.03 & 98.49 \\

(whisper-lfcc-mfcc)-lcnn & 96.93 & 82.50 & 89.09 & 99.22 & 99.88 & 99.55 & 99.12 & 99.14 & 99.10 & 98.87 & 0.04 & 98.25 \\
\midrule

(lfcc)-mesonet & 29.63 & 64.41 & 32.89 & 98.32 & 87.50 & 92.27 & 95.38 & 86.51 & 89.73 & 91.28 & 0.17 & 88.81 \\

(mfcc)-mesonet & 60.49 & 39.10 & 23.95 & 97.29 & 91.61 & 93.90 & 95.71 & 89.36 & 90.90 & 89.88 & 0.18 & 88.43 \\

(lfcc-mfcc)-mesonet & 21.88 & 68.12 & 25.70 & 98.30 & 76.21 & 84.83 & 95.02 & 75.87 & 82.30 & 85.31 & 0.23 & 82.23 \\

(whisper)-mesonet & 39.19 & 92.19 & 46.31 & 98.94 & 64.49 & 66.91 & 96.38 & 65.68 & 66.03 & 95.81 & 0.08 & 84.89 \\

(whisper-lfcc)-mesonet & 56.28 & 63.35 & 43.35 & 98.22 & 76.70 & 81.52 & 96.42 & 76.13 & 79.88 & 87.28 & 0.21 & 82.66 \\

(whisper-mfcc)-mesonet & 40.88 & 67.55 & 44.48 & 98.50 & 91.46 & 94.61 & 96.03 & 90.44 & 92.47 & 92.86 & 0.15 & 90.72 \\

(whisper-lfcc-mfcc)-mesonet & 90.62 & 28.42 & 36.74 & 96.90 & 99.73 & 98.29 & 96.63 & 96.68 & 95.66 & 89.69 & 0.17 & 90.33 \\
\midrule

(lfcc)-specrnet & 87.82 & 63.77 & 73.60 & 98.40 & 99.60 & 98.99 & 97.95 & 98.06 & 97.91 & 97.87 & 0.07 & 96.63 \\

(mfcc)-specrnet & 87.91 & 61.52 & 72.17 & 98.30 & 99.59 & 98.94 & 97.86 & 97.96 & 97.80 & 97.88 & 0.08 & 96.42 \\

(lfcc-mfcc)-specrnet & 92.84 & 57.70 & 70.83 & 98.14 & 99.78 & 98.95 & 97.91 & 97.98 & 97.75 & 97.02 & 0.09 & 95.65 \\

(whisper)-specrnet & 95.33 & 82.19 & 88.14 & 99.21 & 99.82 & 99.51 & 99.04 & 99.06 & 99.02 & 98.81 & 0.04 & 98.06 \\

(whisper-lfcc)-specrnet & 94.20 & 89.00 & 91.48 & 99.51 & 99.75 & 99.63 & 99.28 & 99.29 & 99.28 & 99.24 & 0.03 & 98.71 \\

(whisper-mfcc)-specrnet & 96.77 & 84.41 & 90.07 & 99.31 & 99.87 & 99.59 & 99.20 & 99.21 & 99.18 & 98.27 & 0.04 & 97.97 \\

(whisper-lfcc-mfcc)-specrnet & 87.87 & 90.13 & 88.85 & 99.56 & 99.43 & 99.49 & 99.06 & 99.03 & 99.04 & 99.06 & 0.03 & 98.41 \\

\bottomrule

\end{tabular}%
}
\renewcommand{\arraystretch}{1} 
\end{table*}

\begin{table*}
\centering

\caption{Performance comparison of the baselines on ITW with different feature subsets shown in parentheses. The Avg column is the average of F1-score on the fake class, AUC, and (1-EER). }
\label{tab:itw}
\scalebox{0.91}{
\small
\renewcommand{\arraystretch}{1} 
\begin{tabular}{@{}l|ccccccccccc|cc@{}}
\toprule
 &
  \multicolumn{3}{c}{\cellcolor{RBGreen}Real} &
  \multicolumn{3}{c}{\cellcolor{RBRed}Fake} &
  \multicolumn{6}{c}{\cellcolor{RBBlue}Overall} & \\ \cmidrule(l){3-14} 
\textbf{Model} & 
  P $(\uparrow)$ &
  R $(\uparrow)$ &
  F1 $(\uparrow)$ &
  P $(\uparrow)$ &
  R $(\uparrow)$ &
  F1 $(\uparrow)$ &
  P $(\uparrow)$ &
  R $(\uparrow)$ &
  F1 $(\uparrow)$ &
  AUC $(\uparrow)$ &
  EER  $(\downarrow)$ & 
  Avg. $(\uparrow)$ \\ \midrule

rawnet3 & 99.59 & 99.91 & 99.75 & 99.84 & 99.31 & 99.58 & 99.69 & 99.69 & 99.69 & 99.97 & 0.00 & 99.72 \\ 
\midrule

(lfcc)-lcnn & 99.82 & 99.79 & 99.81 & 99.65 & 99.70 & 99.68 & 99.76 & 99.76 & 99.76 & 99.98 & 0.00 & 99.82 \\ 

(mfcc)-lcnn & 99.78 & 99.77 & 99.78 & 99.62 & 99.63 & 99.63 & 99.72 & 99.72 & 99.72 & 99.99 & 0.00 & 99.77 \\ 

(lfcc-mfcc)-lcnn & 99.73 & 99.75 & 99.74 & 99.58 & 99.55 & 99.56 & 99.68 & 99.67 & 99.67 & 99.99 & 0.00 & 99.74 \\ 

(whisper)-lcnn & 92.88 & 92.45 & 92.66 & 87.34 & 88.02 & 87.68 & 90.82 & 90.80 & 90.81 & 96.54 & 0.10 & 91.49 \\ 

(whisper-lfcc)-lcnn & 99.87 & 99.87 & 99.87 & 99.77 & 99.79 & 99.78 & 99.84 & 99.84 & 99.84 & 99.99 & 0.00 & 99.87 \\ 

(whisper-mfcc)-lcnn & 99.83 & 99.79 & 99.81 & 99.65 & 99.72 & 99.68 & 99.76 & 99.76 & 99.76 & 99.98 & 0.00 & 99.81 \\ 

(whisper-lfcc-mfcc)-lcnn & 99.84 & 99.90 & 99.87 & 99.83 & 99.73 & 99.78 & 99.84 & 99.84 & 99.84 & 99.99 & 0.00 & 99.85 \\ 
\midrule

(lfcc)-mesonet & 99.57 & 99.40 & 99.48 & 98.99 & 99.27 & 99.13 & 99.35 & 99.35 & 99.35 & 99.93 & 0.01 & 99.46 \\ 

(mfcc)-mesonet & 87.27 & 99.55 & 92.33 & 99.17 & 69.47 & 74.91 & 91.69 & 88.37 & 85.86 & 98.82 & 0.03 & 90.23 \\ 

(lfcc-mfcc)-mesonet & 98.73 & 99.30 & 99.01 & 98.81 & 97.84 & 98.32 & 98.76 & 98.76 & 98.76 & 99.86 & 0.01 & 98.97 \\ 

(whisper)-mesonet & 83.97 & 83.35 & 82.75 & 75.37 & 71.11 & 71.05 & 80.78 & 78.80 & 78.40 & 87.68 & 0.20 & 79.68 \\ 

(whisper-lfcc)-mesonet & 93.72 & 99.53 & 96.42 & 99.20 & 87.84 & 92.62 & 95.76 & 95.19 & 95.01 & 99.71 & 0.01 & 96.97 \\ 

(whisper-mfcc)-mesonet & 88.45 & 99.74 & 93.31 & 99.54 & 74.11 & 81.31 & 92.57 & 90.21 & 88.84 & 99.47 & 0.03 & 92.64 \\ 

(whisper-lfcc-mfcc)-mesonet & 97.59 & 98.75 & 98.13 & 97.95 & 95.71 & 96.72 & 97.72 & 97.62 & 97.60 & 99.69 & 0.02 & 98.28 \\ 
\midrule

(lfcc)-specrnet & 99.70 & 99.89 & 99.80 & 99.82 & 99.49 & 99.65 & 99.74 & 99.74 & 99.74 & 99.99 & 0.00 & 99.80 \\ 

(mfcc)-specrnet & 99.63 & 99.83 & 99.73 & 99.72 & 99.38 & 99.55 & 99.66 & 99.66 & 99.66 & 99.97 & 0.00 & 99.70 \\ 

(lfcc-mfcc)-specrnet & 99.82 & 99.82 & 99.82 & 99.69 & 99.69 & 99.69 & 99.77 & 99.77 & 99.77 & 99.99 & 0.00 & 99.82 \\ 

(whisper)-specrnet & 93.00 & 91.83 & 92.40 & 86.51 & 88.29 & 87.37 & 90.59 & 90.51 & 90.53 & 96.49 & 0.10 & 91.38 \\ 

(whisper-lfcc)-specrnet & 99.74 & 99.87 & 99.81 & 99.79 & 99.56 & 99.68 & 99.76 & 99.76 & 99.76 & 99.98 & 0.00 & 99.78 \\ 

(whisper-mfcc)-specrnet & 99.68 & 99.76 & 99.72 & 99.59 & 99.46 & 99.53 & 99.65 & 99.65 & 99.65 & 99.96 & 0.00 & 99.70 \\ 

(whisper-lfcc-mfcc)-specrnet & 99.81 & 99.67 & 99.74 & 99.45 & 99.68 & 99.56 & 99.68 & 99.67 & 99.67 & 99.98 & 0.00 & 99.74 \\

\bottomrule

\end{tabular}%
}
\renewcommand{\arraystretch}{1} 
\end{table*}

\subsection{Cross-Dataset Evaluations}

\begin{figure*}[htbp]
    \centering
    \begin{subfigure}[b]{0.3\textwidth}
        \centering
        \includegraphics[width=\textwidth]{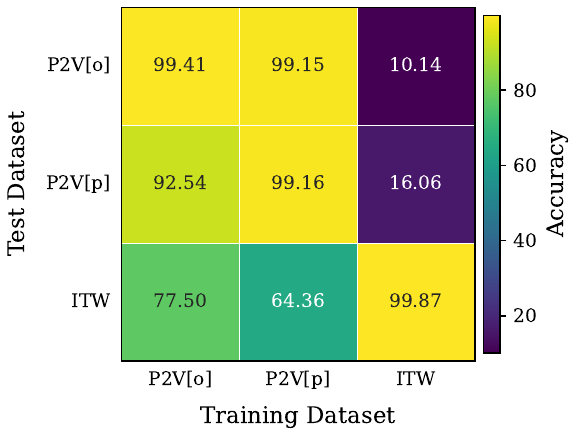}
        \caption{Accuracy}
        \label{fig:cross-dataset-accuracy}
    \end{subfigure}
    \hfill
    \begin{subfigure}[b]{0.3\textwidth}
        \centering
        \includegraphics[width=\textwidth]{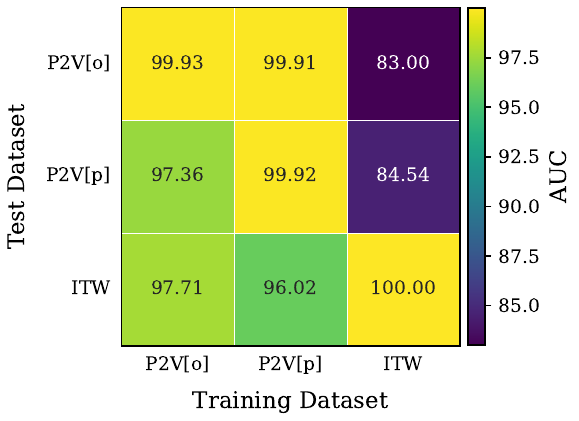}
        \caption{AUC}
        \label{fig:cross-dataset-auc}
    \end{subfigure}
    \hfill
    \begin{subfigure}[b]{0.3\textwidth}
        \centering
        \includegraphics[width=\textwidth]{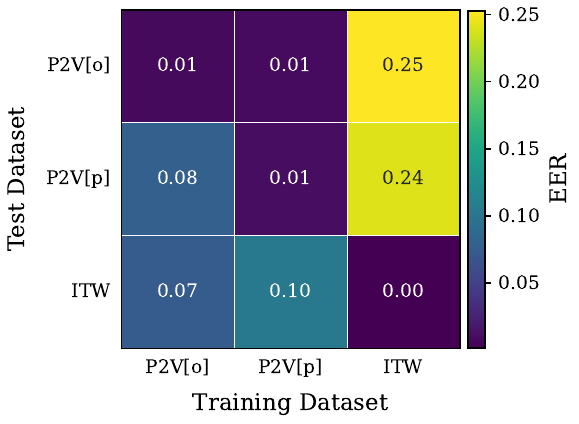}
        \caption{EER}
        \label{fig:cross-dataset-eer}
    \end{subfigure}
    
    \begin{subfigure}[b]{0.3\textwidth}
        \centering
        \includegraphics[width=\textwidth]{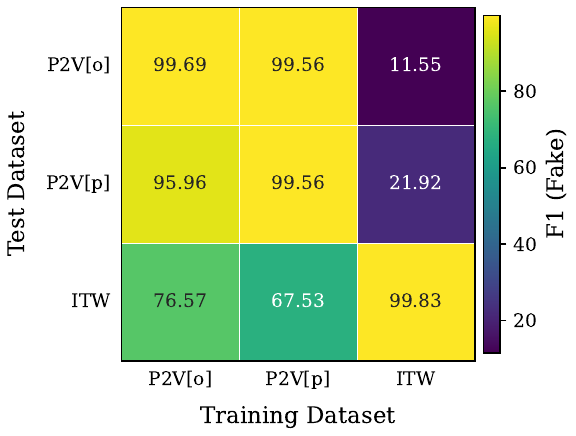}
        \caption{F1 Score (Fake)}
        \label{fig:cross-dataset-f1-fake}
    \end{subfigure}
    \hfill
    \begin{subfigure}[b]{0.3\textwidth}
        \centering
        \includegraphics[width=\textwidth]{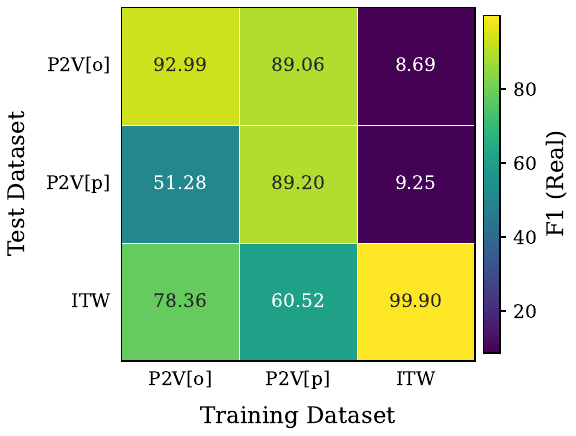}
        \caption{F1 Score (Real)}
        \label{fig:cross-dataset-f1-real}
    \end{subfigure}
    \hfill
    \begin{subfigure}[b]{0.3\textwidth}
        \centering
        \includegraphics[width=\textwidth]{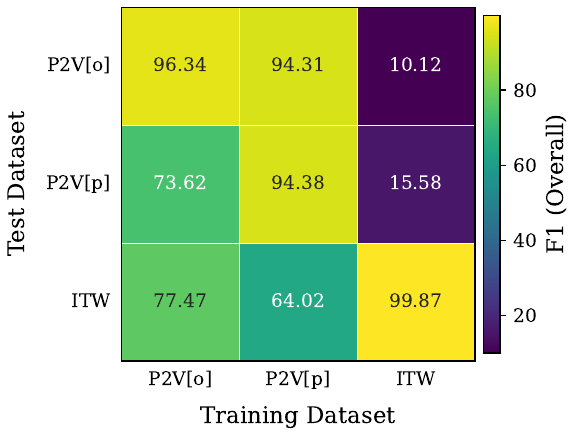}
        \caption{F1 Score (Overall)}
        \label{fig:cross-dataset-f1-overall}
    \end{subfigure}
    
    \begin{subfigure}[b]{0.3\textwidth}
        \centering
        \includegraphics[width=\textwidth]{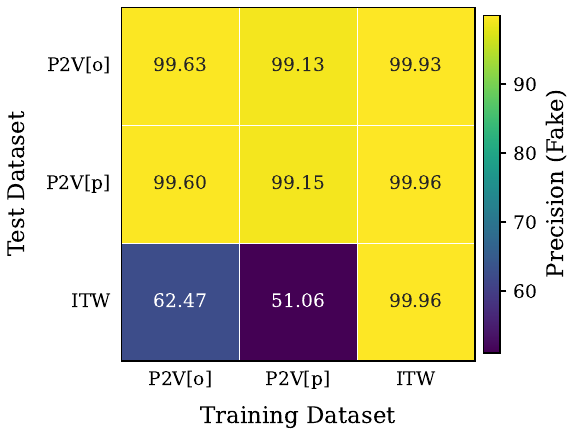}
        \caption{Precision (Fake)}
        \label{fig:cross-dataset-precision-fake}
    \end{subfigure}
    \hfill
    \begin{subfigure}[b]{0.3\textwidth}
        \centering
        \includegraphics[width=\textwidth]{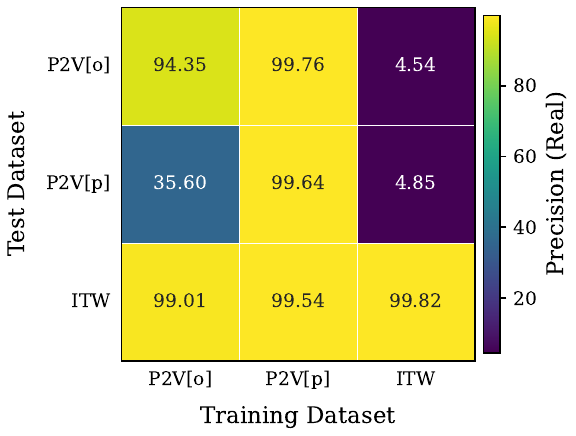}
        \caption{Precision (Real)}
        \label{fig:cross-dataset-precision-real}
    \end{subfigure}
    \hfill
    \begin{subfigure}[b]{0.3\textwidth}
        \centering
        \includegraphics[width=\textwidth]{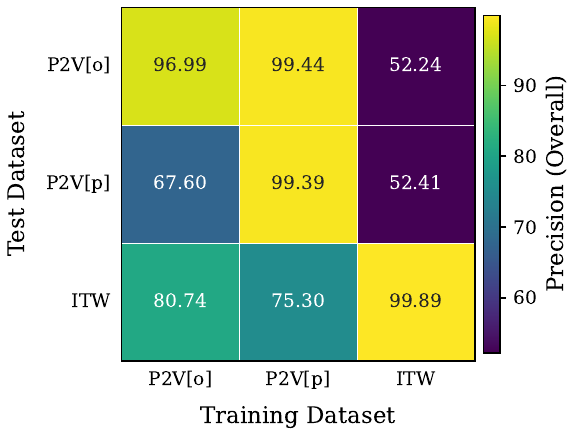}
        \caption{Precision (Overall)}
        \label{fig:cross-dataset-precision-overall}
    \end{subfigure}
    
    \begin{subfigure}[b]{0.3\textwidth}
        \centering
        \includegraphics[width=\textwidth]{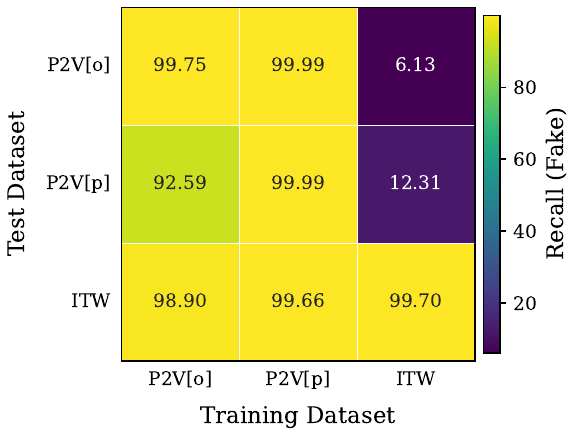}
        \caption{Recall (Fake)}
        \label{fig:cross-dataset-recall-fake}
    \end{subfigure}
    \hfill
    \begin{subfigure}[b]{0.3\textwidth}
        \centering
        \includegraphics[width=\textwidth]{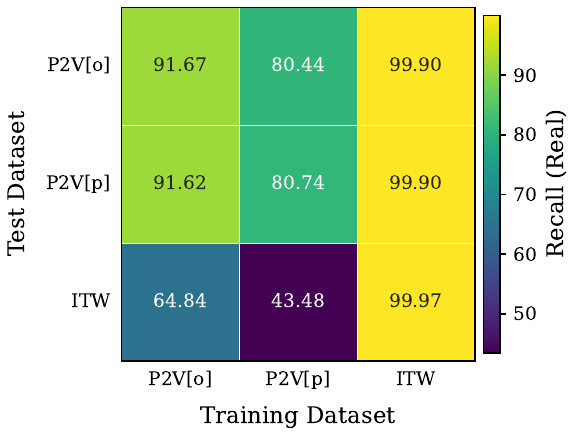}
        \caption{Recall (Real)}
        \label{fig:cross-dataset-recall-real}
    \end{subfigure}
    \hfill
    \begin{subfigure}[b]{0.3\textwidth}
        \centering
        \includegraphics[width=\textwidth]{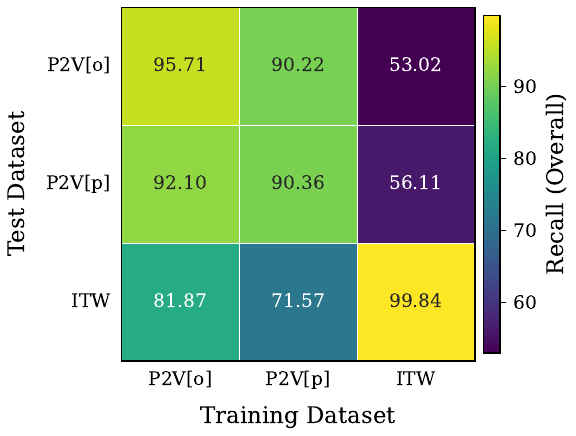}
        \caption{Recall (Overall)}
        \label{fig:cross-dataset-recall-overall}
    \end{subfigure}
    
    \caption{\textbf{Detailed cross-dataset generalization performance metrics.} Each heatmap shows performance for models trained on the dataset indicated on the x-axis and evaluated on the dataset indicated on the y-axis, across 22 baseline models. The metrics include: (a) accuracy, (b) area under the curve (AUC), (c) equal error rate (EER), (d-f) F1 scores for fake class, real class, and overall, (g-i) precision for fake class, real class, and overall, and (j-l) recall for fake class, real class, and overall.}
    \label{fig:cross-dataset-detailed}
\end{figure*}

Figure~\ref{fig:cross-dataset-detailed} reports heatmaps of all the evaluated metrics across 22 baseline models, with rows indicating the test dataset and columns the training dataset. The DDS performance (mean of AUC, F1 score on the fake class, and 1-EER) is reported in the main body.

